\documentclass[aps,prb,twocolumn,a4paper, floatfix]{revtex4}

\usepackage{graphicx}
\begin{document}

\title{A hybrid method for calorimetry with subnanolitre samples using Schottky junctions}

\author{T.~K.~Hakala, J.~J.~Toppari, and P.~T\"orm\"a}

\affiliation{Nanoscience Center, Department of Physics, P.O.Box 35
(YN), FIN-40014 University of Jyv\"askyl\"a, Finland}

\begin{abstract}
A $\mu$m-scale calorimeter realized by using Schottky junctions as a thermometer is presented. Combined with a hybrid experimental 
method, it enables simultaneous time-resolved measurements of variations in both the energy and the heat capacity of subnanolitre 
samples. 
\end{abstract}

\maketitle

\section{INTRODUCTION}

The present micromachining techniques allow the scaling of the calorimeter dimensions down to micrometer or even submicrometer 
scale, which results in high sensitivity and rapid response times. For instance, thermopiles fabricated on a thin SiN membrane 
have been used to determine the catalase activity within a single mouse hepatocyte.\cite{Johannessen} Recently, nanowatt 
sensitivity and time constant of millisecond was achieved by optimizing such a structure.\cite{Chancellor} 
In general, membrane based microcalorimeters using dc-methods have lately been under extensive 
study.\cite{denlinger,olson,Zhang2,cavicchi,torres} 
Additionally, ac-calorimetry methods\cite{Sullivan} have been developed for measurement of heat capacity\cite{Garden,Chateau} and 
thermal conductivity\cite{Zhang} of increasingly small samples.

In this paper we describe the first $\mu$m-scale calorimeter realized by using Schottky junctions as a thermometer. The ability to 
use present IC fabrication techniques makes Schottky junctions a very attractive choice for mass produced, low cost and high 
throughput calorimeters for a variety of applications. Moreover, we utilize a novel measurement method which enables measurements 
of energy changes to be performed simultaneously with ac-calorimetry.\cite{adamovsky} Combined with a short time 
constant ($\sim$20 ms) of the device, it allows direct time-resolved measurements of 
both the variations in the sample heat capacity
and the energy changes due to for example a phase transition or other phenomena. 
The calorimeter may be used in ambient conditions and with liquid samples allowing also real time monitoring, e.g., under a 
microscope, and the operation is near isothermal utilizing only low heating rates, thus making it particularly suitable for 
biological applications.
The performance and reliability of the device and method were
tested. We used subnanolitre drops of DI-water as samples\cite{Chancellor,olson} and measured,
simultaneously, the change of the heat capacity and the latent heat related to the evaporation of the drop.   

\section{EXPERIMENTS}
\subsection{Sample fabrication}

\begin{figure}[htb]
\includegraphics[width=80truemm]{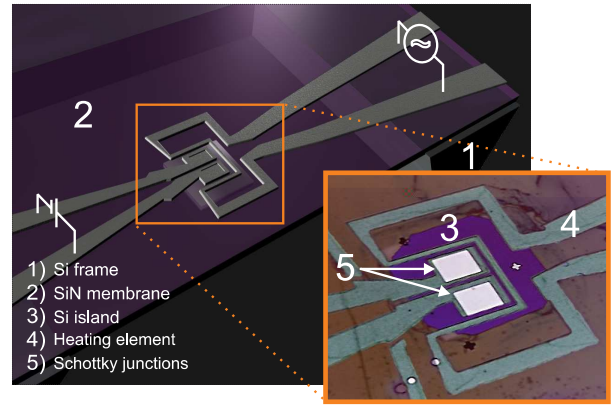}
\caption{A schematic and a micrograph (close-up) of the fabricated sample and the measurement principle.}
\label{fig:Schematic}
\end{figure}

The fabricated calorimeter is composed of two Si/Ti Schottky junctions as a thermometer and a Ti heating element on a small Si 
island supported by a thin (600 nm) SiN-layer (see Fig.~\ref{fig:Schematic}). The samples were fabricated on a lightly boron doped 
(12 $\Omega$cm) Si chip having a SiN layer on both sides. By using photolithography and Reactive Ion Etching (RIE), a square 
opening (1.2$\times$1.2 mm$^2$) was etched on the SiN layer on one side. The opening was then used as a mask for chemical KOH 
etching to form a Si well. The etching was interrupted when there was still about 5--10 $\mu$m of Si at the bottom of the well, 
under which there was the SiN layer of the other side. After that the well was covered with SiN using PECVD deposition. This SiN 
layer was further e-beam patterned and RIE etched by using PMMA as a mask, to leave only a small SiN square at the center of the 
well bottom. A second KOH etching was applied to form a separate Si island onto the SiN membrane. On the other side of that SiN 
membrane, openings on the SiN for the Schottky junctions to the Si island were done by e-beam lithography and RIE. The chip was 
exposed to chemical cleaning \{procedure consisting of 2\% HF (20s)/Piranha (5min)/2\% HF (20s)\}\cite{cleaning} prior to 
deposition of Ti, that acted as a metal for Schottky junctions. Ti was deposited by e-beam evaporator in UHV chamber. Ti layer was 
then patterned and (RIE) etched to form a heating element and the Schottky junctions onto the membrane.

\subsection{Experimental methods}\label{Ch:measetup}

It is well known that Schottky diodes can be used as accurate thermometer at room temperature.
The charge transport through Schottky junctions with low semiconductor doping concentration is dominated by thermionic emission, 
which yields an exponential dependence between the current and the temperature given by $I_\mathrm{tot}=AJ_\mathrm{ST} 
\left[\exp(qV/k_\mathrm{B}T)-1\right]$.\cite{Sze} Here $V$ is the voltage across the junction, $A$ the area of the junction, $q$ 
the electron charge, $k_\mathrm{B}$ the Boltzmann constant and $T$ the absolute temperature. The saturation current density 
$J_\mathrm{ST}$ is defined by $J_\mathrm{ST}\equiv A^*T^2\exp(-q\phi_{B}/k_\mathrm{B}T)$, where $A^*$ is the effective Richardson 
constant and $\phi_{B}$ the barrier height. We consider temperature changes $\Delta T$ small enough so that the dependence of the 
current on temperature, $I(T)$, can be linearized, i.e.\ $I(T)\propto (dI/dT)\Delta T$. In the present configuration there are two 
Schottky junctions in series, one being always forward biased and the other reverse biased. The zero bias resistance of the 
fabricated  samples with 35$\times$45 $\mu$m$^2$ Schottky contacts in ambient conditions is 5-10 M$\Omega$ and the variation of 
resistance about 10 \%/K.

At the equilibrium, the temperature of the Si island $T$ is governed by the net power $P_\mathrm{in}$ applied to it, and by the 
heat conductance $K$ [W/K] of the SiN membrane to a heat bath formed by the Si chip at temperature $T_0$. In general, one can 
write 
\begin {equation}
T=T_0+P_\mathrm{in}/K, 
\label{equi}
\end{equation}
if the variations in $P_\mathrm{in}$ are slower than time constant of the calorimeter $\tau=(C+\Delta C)/K$. Here $C+\Delta C$ is 
the total heat capacity of the Si island ($C$) and possible sample on it ($\Delta C$). 
An ac voltage $V=V_0\sin(\omega t)$ applied via the Ti-heater, i.e., the power $P_\mathrm{H}=V^2_0/R\times sin^2(\omega t)$, where 
$R$ is the resistance of the heating element, produces by Joule heating a temperature fluctuation of the sample with a frequency 
$2 \omega$. If $\omega\sim\tau^{-1}$, the equilibrium result does not apply anymore and one obtains for the temperature of the 
island
\begin{equation}
T\!=\!T_0\!+\!\frac{V_0^2}{2RK}\!\!\left[\!1\!-\!\frac{e^{-t/\tau}}{1\!+
\!(2\omega\tau)^{\!-2}}\!
-\! \frac{\cos(2\omega t\!+\!\varphi)}{\sqrt{1\!+\!(2\omega\tau)^2}} \right]\!\! 
+\!\frac{P_\mathrm{S}}{K}\!,
\label{nonequi}
\end{equation}
where $\varphi=\arctan(2\omega\tau)$ and $P_\mathrm{S}$ is the power produced by the sample (e.g.\ due to a chemical or biological 
reaction), or by some external source. Consequently, the dc-biased Schottky junctions will carry a current, directly proportional 
to $T$ above, with one component having the frequency of $2\omega$ and another being the dc component proportional to 
$P_\mathrm{S}$ and the constant rms power fed in via the heater, $V_0^2/2RK$.  

\begin{figure}[htb]
\includegraphics[width=80truemm]{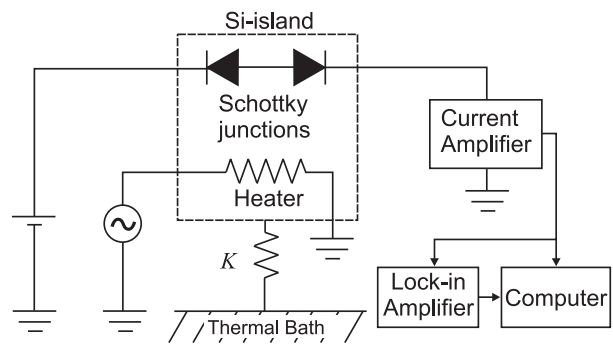}
\caption{A schematic of the measurement setup.}
\label{fig:Measchematic}
\end{figure}

Our calorimeter function is based upon the combination of two calorimeter operation modes: the heat conduction mode, which 
measures $P_\mathrm{S}/K$ via the dc component, and the ac-calorimeter mode with an ac voltage of frequency $\omega\sim\tau^{-1}$ 
applied to the heater (see Figs. \ref{fig:Schematic} and \ref{fig:Measchematic}). According to Eq.~(2), the amplitude of the 
produced $2\omega$ temperature component depends on the heat capacity of the sample as
\begin{equation}
\delta T=\frac{V^2}{2R}\frac{1}{\sqrt{K^2+(2\omega)^2(C+\Delta C)^2}},
\label{eq:dT}
\end{equation}
where the term $\Delta C$ describes the heat capacity of the sample which can also slowly vary in time, e.g., due to evaporation 
of a liquid sample drop. Thus, by simultaneous measurement of the dc and ac signals, one is able to detect both the variation in 
temperature and the variation in heat capacity, respectively. 

The simultaneous monitoring of the dc and ac signals was performed by using a measurement setup shown in 
Fig.~\ref{fig:Measchematic}. The dc-bias of the Schottky junctions was set to 60 mV and the sinusoidal signal for the heating 
element had rms power of 37 $\mu$W and a frequency of 10 Hz corresponding to the situation $\omega\sim\tau^{-1}$. The dc component 
of the temperature sensitive current through the Schottky junctions  was measured directly using DL Instruments low noise voltage 
preamplifier 1201 and to measure the 20 Hz ac component, Stanford Research SR830 digital lock-in amplifier was used. These both 
signals were recorded by a computer with a sampling rate of 1 kHz. The fluctuating ac signal component was extracted from the 
measured dc signal by digital notch filtering.

\subsection{Sample delivery}\label{Ch:pipetting}

\begin{figure}[htb]
\includegraphics[width=80truemm]{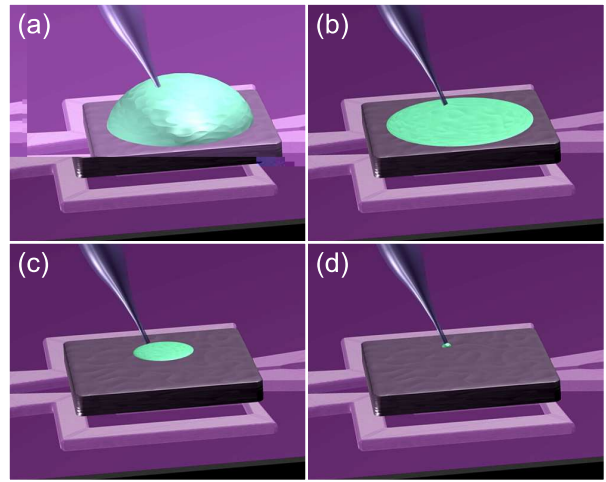}
\caption{(a) -- (d) Cartoon of evaporation of a water drop after injection. In the end, a small amount of water stays to bridge 
the end of the pipette and the Si island of the calorimeter. This configuration, shown in (d), is considered as the "empty 
device". The thermal conductivity $K$ and the heat capacity $C$ are determined for this "empty device", thus accurately and 
systematically including the effect of the pipette.}
\label{fig:Pipetting}
\end{figure}

The sample to be measured, i.e., drop of DI-water is delivered to the Si island of the calorimeter constructed onto the center of 
the SiN membrane (see Figs.~\ref{fig:Schematic} and \ref{fig:Pipetting}) using a very narrow pulled glass pipette. The injection 
is carried out by having the pipette back end attached to a pressure device (ASI MPPI-2 microinjector) which can supply a short 
pressure pulse with adjustable pressure and pulse time. In principle, this device allows one to precisely and reproducibly (from 
measurement to measurement) control the amount of liquid injected. However, it is not possible to quantify the amount since the 
total amount depends on the viscosity of the liquid, the surface effects between the liquid and the pipette, the pipette 
dimensions, etc.

All the time during the sample delivery and the measurement, the end of the pipette is at the very near proximity of the Si island 
of the calorimeter. This is necessary since, due to high surface tension of the water, the small drops used in this experiment 
will not have sufficient mass to detach from the tip of the pipette unless they touch the Si island surface. During the pressure 
pulse, the drop size increases until it touches the hydrophilic surface of the Si island and spreads on. After the pulse, the drop 
size starts to decrease due to evaporation of the drop and also due to the reflux of the water back to the injector as will be 
discussed later.

After the evaporation of the drop, a small amount of water is left to thermally bridge the pipette and the Si island (see 
Fig.~\ref{fig:Pipetting}). This is due to hydrophilicity of the surface of the Si island and the surface tension of water, and it 
creates an additional pathway for heat transfer between the calorimeter and the environment. Since the relative positioning of the 
tip of the pipette and the calorimeter remains constant throughout the whole experiment, it is expected that this additional 
pathway for heat transfer will simply increase the total heat conductivity $K$ in our simple model by a constant factor which can 
be experimentally determined. Thus the total thermal conductivity may be assumed to be constant over the time span of the 
experiment.

Prior to measurement of the device parameters (see next section), a drop of water was injected onto the Si island surface which 
was then allowed to evaporate to form the water bridge between the calorimeter and the pipette as described above. This 
configuration is now considered to be the empty device, since after each injection the system returns to this state. The pipette 
is kept on its place during all the further experiments. Thus, the obtained parameter values already include the effect of the 
additional heat conductivity through the water, i.e., the effect of the pipette slightly touching the device. Importantly, this 
effect can be accurately measured, and it remains constant through the whole measurement process, which makes it controllable. 
However, for more practical use of the device, other delivery systems, e.g., microfluidistic channels, could be implemented.

\subsection{Calibration}

The dc signal (with 60 mV dc-bias) dependence on the temperature was calibrated by using a Pt-100 thermometer as a reference, 
resulting in a linear relation with a slope of $9.5\times 10^{-10}$ A/K between the current and the temperature near the operation 
temperature $23^{\circ}$C. Additionally, the relaxation method was used to determine the thermal conductance, $K$, and heat 
capacity, $C$, of the empty device (including the pipette for water, see above and Fig.~\ref{fig:Pipetting})
without the sample: To obtain the thermal conductance according to the Eq.~(\ref{equi}), we measured the dc current as a function 
of slowly increasing dc power (applied to the heating element), yielding $6.95 \times 10^{-5}$ W/K. The time constant of the 
calorimeter was determined  by measuring the dc signal response to the step pulse fed into the heater, resulting in a time 
constant of 21 ms. Finally, the calorimeter heat capacity was obtained as the product of the time constant and the conductance 
which yielded a value of $1.53\times 10^{-6}$ J/K.

\section{RESULTS AND DISCUSSION}

\subsection{Measured signals for subnanolitre drops of water}

\begin{figure}[htb]
\includegraphics[width=80truemm]{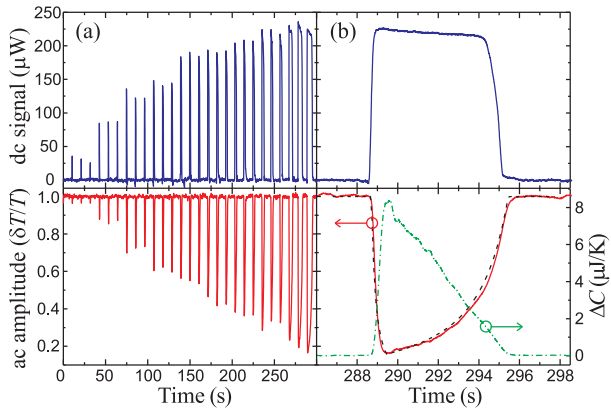}
\caption{(a) The dc (blue) and ac (red) signals of the thermometer for several consequent injections, the injection times ranging 
from 10 ms up to 60 ms. (b) A closeup of both signals with 60 ms injection time. The dash-dotted (green) line is the heat capacity 
of the drop, $\Delta C$, obtained from the ac signal as a function of time and the dashed line (black) shows the theoretical 
signal calculated along Eq.~(3) assuming linear decrease in $\Delta C$. The two arrows show the corresponding axes for the 
curves.}
\label{fig:Signals}
\end{figure}

The validity of our measurement method was evaluated by injecting several subnanolitre-scale drops of DI-water with different 
(undetermined) volumes onto the calorimeter, and by measuring the ac and dc responses as functions of time as explained in section 
\ref{Ch:measetup}. In Fig.~\ref{fig:Signals}a) is shown the dc and ac signals of the thermometer with several injection times of 
the nanoinjector, ranging from 10 ms up to 60 ms (injection pressure being constant $\sim0.5$ Bar). During all the measurements, 
the pipette was in thermal contact with the device as described above in section \ref{Ch:pipetting}.  

The figure \ref{fig:Signals}b) illustrates the function of the sensor in detail for 60 ms injection time. From the figure one can 
see the abrupt change in the two signals when the injection of drop takes place, i.e., at around 288.5 s. The dc signal increases, 
indicating the temperature decrease due to evaporation of the drop while the ac signal decreases due to increase in heat capacity 
of the system. Between the time 289--295.5 s the dc signal remains approximately constant, which implies that the power 
consumption due to evaporation ($P_\mathrm{S}$ in Eq.~(\ref{nonequi})) and thus the surface area of the drop, remains almost 
constant. This is due to hydrophilicity of the Si island surface, which causes a flat shape of the drop and the height of it to 
vanish almost completely before the horizontal area starts to decrease during evaporation (see Fig.~\ref{fig:Pipetting}). This 
also implies that the volume of the evaporated part of the drop is decreasing with approximately constant rate. During this 
timeframe the ac signal is increasing which is due to continuous decrease in heat capacity, $\Delta C$, of the system as the drop 
evaporates, which is also shown as a dash-dotted (green) curve in Fig.~\ref{fig:Signals}b). As the heat capacity is proportional 
to volume, this is consistent with the conclusions made from the dc signal.

In figure \ref{fig:Signals}b) the dashed line shows the calculated (along Eq.~(3)) variation of ac signal amplitude as a function 
of time assuming a constant reducing rate of volume after the injection. Also, a finite time for injection, during which the drop 
is
growing linearly, is assumed in calculation. These assumptions can be clearly verified from the  curve showing $\Delta C$ as a 
function of time. Heat capacity of the drop, $\Delta C$, as well as the heat capacity, $C$, and conductance, $K$, of the 
calorimeter for the calculation are obtained from the measurement and calibration data. The excellent agreement of the calculated 
curve with the measured one further verifies that the reducing rate of the total heat capacity of the sample during the 
evaporation is approximately constant in time, as the dc signal already suggested for the evaporated part. Since the total 
reduction of the heat capacity is due to evaporation and also due to the reflux of the injected water back to the pipette, as will 
be further explained in next section, this implies that also the reflux happens at a constant rate. 

\subsection{Detection of nonidealities in sample delivery}\label{ch:reflux}

In general, the heat capacity, $\Delta C$ of the sample can be expressed in terms of specific heat $c$ and mass $m$ of the sample 
as $\Delta C\!=\!cm$. Furthermore, for samples that evaporate, $\Delta C\!=\!cm\!=\!c E/\ell$, where $E$ is the total energy 
required for evaporation of a drop and $\ell$ is the latent heat of evaporation. From the measured dc signal one is able to obtain 
the energy $E$ as the time integral of the signal, and the variation of the heat capacity of the system is related to the ac 
signal via Eq.~(\ref{eq:dT}). Such complementary information allows to understand details of the process and eliminate the effect 
of measurement non-idealities (which are of increasing importance when the sample volume decreases), as will be shown in the 
following. Furthermore, for evaporating samples, it offers a test of consistency since the heat capacity can be determined in two 
ways: from the ac signal, and, in addition, from the dc signal via the relation $\Delta C\!=\!c E/\ell$.

As the main non-ideality in the measurement, it was observed that the non-ideal backpressure applied to the nanoinjector produced 
a reflux of the injected drop back into the pipette
during evaporation (the backpressure is used to compensate the capillary forces that result in a reflux of the injected liquid 
back to the pipette). Therefore the energy of evaporation $E$ obtained by integrating the dc signal does not give, without a 
correction, the initial heat capacity of the sample from $\Delta C\!=\!c E/\ell$. The required correction was obtained by 
performing measurements for drops of different (unknown) volumes until they were fully evaporated, and by 1) determining the mass 
of the initial drop from the ac signal as the ratio $m_\mathrm{C}=\Delta C/c$ of the maximum variation of heat capacity $\Delta C$ 
and the specific heat of water $c$, 2) determining the mass of the evaporated part of drop from the dc signal as the ratio 
$m_\mathrm{E}=E/\ell$ of the energy required for evaporation $E$ and the latent heat of water $\ell$. The difference between the 
mass of the initial drop and that of the evaporated part then gives the mass of the part that was refluxed to the injector.

\begin{figure}[htb]
\includegraphics[width=80truemm]{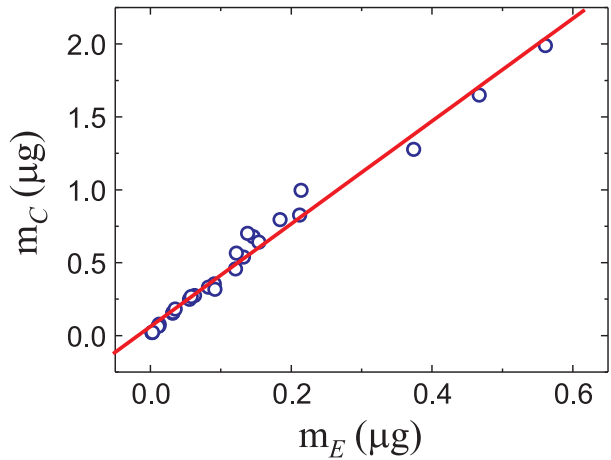}
\caption{Relation between the initial mass of the drop $m_C$ and the mass of the evaporated part $m_E$. Circles correspond to the 
measured data and line is a linear fit to it.}
\label{fig:refluxfit}
\end{figure}

A plotting of $m_\mathrm{C}$ against $m_\mathrm{E}$ for each drop, as shown in Fig.~\ref{fig:refluxfit}, reveals that the mass 
$m_\mathrm{E}$ obtained from evaporation energy is systematically smaller than the mass $m_\mathrm{C}$ obtained from minimum of 
the ac signal. This result is easy to understand by noting that since the pipette sucks some of the liquid back during the 
evaporation, the mass $m_\mathrm{E}$ obtained from evaporation energy must be smaller than the mass $m_\mathrm{C}$ obtained from 
the initial heat capacity of the drop (ac signal). Furthermore, since the reflux is due to a constant pressure difference, it is 
supposed to happen at approximately constant rate, which can also be verified from the signals in Fig.~\ref{fig:Signals} as 
explained in previous section \ref{ch:reflux}. As shown by the Fig.~\ref{fig:refluxfit}, as much as 72\% of the initial mass is 
lost by the reflux in every pipetting. 

Such measurements can also be used to calibrate the pipette for different liquids, surfaces, etc., which is often found to be a 
problem in nano- and subnanolitre injections. Therefore, the hybrid method and the device we use allow to accurately distinguish 
various non-idealities of the measurement from the chemical/physical/biological phenomena of interest. This is certainly a 
strength, especially at nanoscale where non-idealities in the measurements are difficult to avoid.

\subsection{Evaluation of the calorimeter}

\begin{figure}[htb]
\includegraphics[width=80truemm]{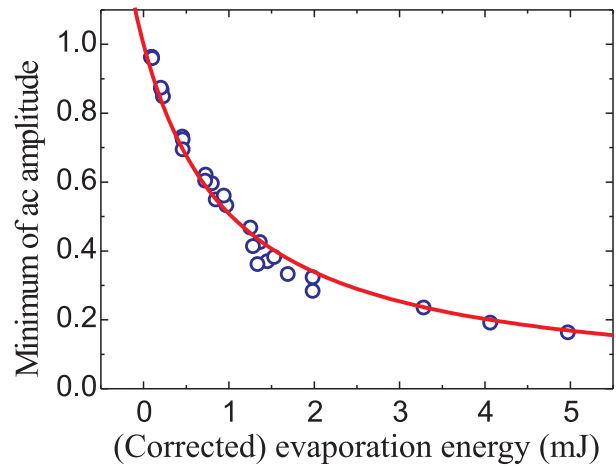}
\caption{The measured minimum values of the normalized ac signal and the measured total energy required for drop evaporation (with 
the reflux correction), together with the theoretical fit obtained using the relation of evaporation energy and heat capacity and 
Eq.(\ref{eq:dT}).}
\label{fig:fitting}
\end{figure}

Finally, as a test of performance and consistency of our method, we present in figure \ref{fig:fitting} the following analysis: We 
performed measurements for different sizes of the drops. For each drop, the information given by the two signals, ac and dc, is 
plotted as a point in a two dimensional graph where one axis is the minimum of the normalized ac lock-in signal (corresponds to 
the heat capacity of the initial sample via Eq.~(\ref{eq:dT})), and the other axis is the integrated dc signal giving the 
evaporation energy $E$. The latter yields the mass of the evaporated part of the drop $m_\mathrm{E}$ via the relation 
$m_\mathrm{E} = E/\ell$. However, we cannot use the mass $m_\mathrm{E}$ as such since it excludes the part gone back to the 
pipette by the reflux, and when considering e.g. the minimum value of the ac signal, one has to use the mass of the initial drop, 
which includes the part that later will be gone back to the pipette by the reflux. From the analysis presented above, we know the 
ratio between the evaporated and the initial masses to be constant, $\sim 0.28$. Thereby we can use a corrected value of mass, 
i.e. $m = m_\mathrm{C}\approx m_\mathrm{E}/0.28 = E/(0.28\,\ell)$ and thus corrected energy $E_\mathrm{C} \equiv E/0.28$, which is 
plotted as $x$-axis in \ref{fig:fitting}.  

As a fit to the theory and test of consistency, we then take the measured values of $E$, use the relation $\Delta C= cm = 
(c/\ell)E_\mathrm{C}$ (with $c/\ell$ as the only fitting parameter), insert this value of $\Delta C$ in Eq.~(\ref{eq:dT}), and 
plot the resulting curve in the figure. It corresponds excellently to the measured data points. The values for $C$ and $K$ used in 
this analysis are taken from the measured calibration data. From the value given for $c/\ell$ by the fitting and using the known 
value for $c$ or $\ell$, we obtain the correct value for $\ell$ or $c$, respectively, with the accuracy of about 7\%. This 
accuracy characterizes the overall performance of the device and method at this stage, and can be improved by optimizing the 
sample fabrication and the measurement setup. Note that for evaporating samples where one of the parameters $c$, $\ell$ or volume 
is known, the method can used for determining the two others. 

The ac signal also allows high accuracy mass measurements: the smallest measured change of heat capacity (for 10 ms injection 
time) was approximately $10^{-7}$ J/K which, for water, corresponds to 20 ng mass. This could be extremely useful for example when 
measuring a reaction heat of some chemical or biological reaction, since one is able to simultaneously measure the amounts of the 
reagents added and the energy produced or consumed by the reaction. 

\section{CONCLUSIONS}

In summary, we have developed a microcalorimeter capable of providing simultaneous quantitative time-resolved information of the 
sample temperature, power and total energy production/consumption as well as of sample heat capacity and its variations. The small 
heat capacity of the device which enables fast and sensitive measurements carried out in room temperature, together with the mass 
production potential of the Schottky junctions, hold a promise for high throughput and cost effective tool for biological and 
chemical applications. 

\section*{ACKNOWLEDGMENTS}
We thank Academy of Finland and EUROHORCs (EURYI award,
Academy project number 205470).


\begin{thebibliography}{99}
\footnotesize

\bibitem{Johannessen} E.~A.~Johannessen, J.~M.~R.~Weaver, P.~H.~Cobbold, and J.~M.~Cooper,
Appl.~Phys.~Lett. {\bf 80}, 2029 (2002).

\bibitem{Chancellor} E.~B.~Chancellor, J.~P.~Wikswo, F.~Baudenbacher, M.~Radparvar, and D.~Osterman,
Appl.~Phys.~Lett. {\bf 85}, 2408 (2004).

\bibitem{denlinger} D.~W.~Denlinger, E.~N.~Abarra, K.~Allen, P.~W.~Rooney, M.~T.~Messer, S.~K.~Watson, and F.~Hellman, 
Rev.~Sci.~Instrum. {\bf 65}, 946 (1994).

\bibitem{olson} E.~A.~Olson, M.~Y.~Efremov, A.~T.~Kwan, S.~Lai, V.~Petrova, F.~Schiettekatte, J.~T.~Warren, M.~Zhang, and 
L.~H.~Allen, Appl.~Phys.~Lett. {\bf 77}, 2671 (2000).

\bibitem{Zhang2} Y.~Zhang and S.~Tadigadapa,
Appl.~Phys.~Lett. {\bf 86}, 034101 (2005).

\bibitem{cavicchi} R.~E.~Cavicchi, G.~E.~Poirier, N.~H.~Tea, M.~Afridi, D.~Berning, A.~Hefner, J.~Suehle, M.~Gaitan, S.~Semancik, 
and C.~Montgomery, Sensors \& Actuators {\bf 97}, 22 (2004).

\bibitem{torres} F.E Torres, {\it et al.} Proc.~Nat.~Acad.~Sci. {\bf 101}, 9517 (2004).

\bibitem{Sullivan} P.~F.~Sullivan and G.~Seidel, Phys.~Rev. {\bf 173}, 679 (1968).

\bibitem{Garden} J.-L.~Garden, E.~Ch\^ateau, and J.~Chaussy,
Appl.~Phys.~Lett. {\bf 84}, 3597 (2004).

\bibitem{Chateau} E.~Ch\^ateau, J.-L.~Garden, O.~Bourgeois, and J.~Chaussy,
Appl.~Phys.~Lett. {\bf 86}, 151913 (2005).

\bibitem{Zhang} M.~Zhang, M.~Yu.~Efremov, E.~A.~Olson, Z.~S.~Zhang, and L.~H.~Allen,
Appl.~Phys.~Lett. {\bf 81}, 3801 (2002).

\bibitem{adamovsky} Similar measurements, but for polymer samples and using thermopiles,
have been performed in S.~Adamovsky and C.~Schick,
Thermochim.~Acta {\bf 415}, 1 (2004).

%

\bibitem{cleaning} T.~Kuroda, Z.~Lin, H.~Iwakuro, and S.~Sato, J.~Vac.~Sci.~Technol.~B {\bf 15}, 232 (1997).

\bibitem{Sze} S.~M.~Sze, {\it Physics of Semicondutor Devices} (Wiley, New York, 1981).

\end{thebibliography}
\end{document}